\documentclass{article}
\usepackage[final]{graphics}
\usepackage{amsfonts,amsbsy}
\font\tenmsa=msam10
\font\sevenmsa=msam7
\font\fivemsa=msam5
\font\tenmsb=msbm10
\font\sevenmsb=msbm7
\font\fivemsb=msbm5
\newfam\msafam
\newfam\msbfam
\textfont\msafam=\tenmsa  \scriptfont\msafam=\sevenmsa
\scriptscriptfont\msafam=\fivemsa
\textfont\msbfam=\tenmsb  \scriptfont\msbfam=\sevenmsb
\scriptscriptfont\msbfam=\fivemsb

\def\empile#1\over#2{\mathrel{\mathop{\kern 0pt#1}\limits_{#2}}}

\catcode`\@=11

\newcount\@tempcntc
\def\@citex[#1]#2{\if@filesw\immediate\write\@auxout{\string\citation{#2}}\fi
  \@tempcnta\z@\@tempcntb\m@ne\def\@citea{}\@cite{%
        \@for\@citeb:=#2\do%
    {\@ifundefined{b@\@citeb}%
        {\@citeo\@tempcntb\m@ne\@citea%
                \def\@citea{,\penalty\@m\ }{\bf ?}\@warning%
                {Citation `\@citeb' on page \thepage \space undefined}}%
        {\setbox\z@\hbox{\global\@tempcntc0\csname b@\@citeb\endcsname\relax}
     \ifnum\@tempcntc=\z@ \@citeo\@tempcntb\m@ne%
       \@citea\def\@citea{,\penalty\@m}%
       \hbox{\csname b@\@citeb\endcsname}%
     \else%
      \advance\@tempcntb\@ne%
      \ifnum\@tempcntb=\@tempcntc%
      \else\advance\@tempcntb\m@ne\@citeo%
      \@tempcnta\@tempcntc\@tempcntb\@tempcntc\fi\fi}}\@citeo}{#1}}%

\def\@citeo{\ifnum\@tempcnta>\@tempcntb\else\@citea
  \def\@citea{,\penalty\@m}%
  \ifnum\@tempcnta=\@tempcntb\the\@tempcnta\else
   {\advance\@tempcnta\@ne\ifnum\@tempcnta=\@tempcntb \else
\def\@citea{--}\fi
    \advance\@tempcnta\m@ne\the\@tempcnta\@citea\the\@tempcntb}\fi\fi}

\catcode`\@=12

\global\mathchardef\lesssim "142E

\newcommand{\slL}{\raise.15ex\hbox{$/$}\kern-.53em\hbox{$L$}}
\newcommand{\slP}{\raise.15ex\hbox{$/$}\kern-.53em\hbox{$P$}}
\newcommand{\slp}{\raise.1ex\hbox{$/$}\kern-.63em\hbox{$p$}}
\newcommand{\slq}{\raise.1ex\hbox{$/$}\kern-.63em\hbox{$q$}}
\newcommand{\slv}{\raise.1ex\hbox{$/$}\kern-.63em\hbox{$v$}}
\newcommand{\slR}{\raise.15ex\hbox{$/$}\kern-.53em\hbox{$R$}}
\newcommand{\slQ}{\raise.15ex\hbox{$/$}\kern-.53em\hbox{$Q$}}
\newcommand{\slK}{\raise.15ex\hbox{$/$}\kern-.53em\hbox{$K$}}
\newcommand{\slk}{\raise.15ex\hbox{$/$}\kern-.53em\hbox{$k$}}
\newcommand{\slSigma}{\raise.15ex\hbox{$/$}\kern-.53em\hbox{$\Sigma$}}
\newcommand{\slcalP}{\raise.15ex\hbox{$/$}\kern-.63em\hbox{$\cal P$}}
\newcommand{\slA}{\raise.15ex\hbox{$/$}\kern-.73em\hbox{$A$}}
\newcommand{\slbfA}{\raise.15ex\hbox{$/$}\kern-.73em\hbox{${\imb A}$}}
\newcommand{\slpartial}{\raise.15ex\hbox{$/$}\kern-.53em\hbox{$\partial$}}
\newcommand{\slcalD}{\raise.15ex\hbox{$/$}\kern-.73em\hbox{$\cal D$}}

\newcommand{\be}{\begin{equation}}
\newcommand{\ee}{\end{equation}}
\newcommand{\bea}{\begin{eqnarray}}
\newcommand{\ena}{\end{eqnarray}}

\def\build#1\over#2{\mathrel{\mathop{\kern 0pt#1}\limits_{#2}}}

\newcount\toto
\def\strip[hep-ph/0204#1]{#1}
\def\addoneto#1{\toto=#1\relax%
        \global\advance\toto by 1\relax%
        \the\toto}
\def\subtractoneto#1{\toto=#1\relax%
        \global\advance\toto by -1\relax%
        \the\toto}

\font\tenimbf=cmmib10 at 10pt
\font\sevenimbf=cmmib10 at 7pt
\font\fiveimbf=cmmib10 at 5pt
\newfam\imbf
\textfont\imbf=\tenimbf
\scriptfont\imbf=\sevenimbf
\scriptscriptfont\imbf=\fiveimbf
\def\imb{\fam\imbf\tenimbf}

\begin{document}
\title{\bf{2PI effective action and gauge dependence identities}}

\author{M.E.Carrington\footnote{Electronic address:
{\tt carrington@brandonu.ca}} \\ Department of Physics, Brandon University, Manitoba, R7A 6A9 Canada\and G.Kunstatter\footnote{Electronic address: {\tt g.kunstatter@uwinnipeg.ca}}\\ Department of Physics, University of Winnipeg, Winnipeg, Manitoba, R73 2E9 Canada \and H. Zaraket\footnote{Electronic address: {\tt zaraket@theory.uwinnipeg.ca}} \\ Department of Physics University of Winnipeg, Winnipeg, Manitoba, R73 2E9 Canada}

\maketitle

\begin{abstract}
The problem of maintaining gauge invariance when truncating the two particle 
irreducible  (2PI) effective action has been studied recently by several authors. Here we give a simple and very general derivation of the 
gauge dependence identities for the off-shell 2PI effective action. We consider the case where the gauge is fixed by an
arbitrary function of the quantum gauge field, subject only to the restriction that the Faddeev-Popov matrix is invertable. We also study the background field gauge. We address the role that these identies play in solving gauge invariance problems associated with physical quantities calculated using a truncated on-shell 2PI effective action.
\end{abstract}
\vskip 4mm
\centerline{\hfill }

\section{Introduction}
Considerable progress has been achieved in the study of the dynamics of quantum fields in and 
out of equilibrium. Collective effects and long range interactions are intrinsic to deconfined QCD matter in heavy ion collisions. At 
equilibrium, the hard thermal loop effective theory is the appropriate gauge invariant theory  
to describe most collective and long range effects. On the other hand, kinetic 
theories provide the best method to describe a near equilibrium situation in the small coupling 
constant regime. Far 
from equilibrium and/or at large coupling constant other techniques need to be developed. 
A promising candidate is the 2PI effective action method. Unfortunately, practical calculations 
necessitate the use of approximate versions of the exact 2PI effective action. The approximated, or 
truncated, 2PI effective action is simply a 
Schwinger-Dyson resummation of the two point function. Without further resummation for 
the vertex functions, this resummation necessarily leads to gauge 
dependent results for most physical quantities.

The study of the gauge dependence of functional methods in quantum field theory has a long history. One of the first contributions was made by Nielsen \cite{Nielsen} who showed that the explicit dependence of the 1PI effective potential on the gauge parameter is compensated for by the gauge parameter dependence of the expectation value, and that the effective potential is gauge parameter independent. In \cite{KobesKR1,KobesKR2} a general functional formalism was used to derived the gauge dependence of the associated n-point functions. These results were then used to address the gauge fixing dependence of the one loop  QCD plasma damping rate. 

  In this paper we derive an expression for the gauge fixing dependence of the exact, off-shell 2PI effective action for any gauge fixing function that has an invertible Faddeev-Popov matrix. Using this
expression for the exact effective action we will analyze the 
gauge dependence of the truncated effective action and show that the gauge dependence always occurs at higher order, within any self-consistent truncation scheme. This verifies the expectation that, in complete analogy with the 1PI formalism, gauge invariance problems that occur in specific calculations using the 2PI formalism should ultimately be traceable to an inconsistency in the approximation scheme. In addition, we will show that the method 
we use is easily generalized to the case of background field gauges. The  
background field gauge is of interest because of the recent results of Mottola \cite{Motto1}. Mottola 
suggested a modified form of the 2PI effective action in which Ward identities for background and 
quantum gauge fields are both satisfied, under certain conditions.
We note that the leading order gauge invariance of the truncated effective action has already been obtained by a different method in \cite{ArrizS1}.

The paper is organized as follows. In section (\ref{1PI}) we discuss gauge dependence in the
context of the 1PI effective action. In section (\ref{general}) we review the 2PI formalism. In 
section (\ref{nielsen}) we derive 
the 2PI Nielson identities. In section (\ref{Truncation}) we discuss the circumstances under which truncated 2PI equations can be expected to lead to gauge dependent results for physical quantities. In section (\ref{backg}) we look at the background field gauge and in section (\ref{conc}) we present some 
conclusions.  The connection between our result and that 
of \cite{ArrizS1} is discussed in the appendix. 

Throughout this paper we use the compact notation of DeWitt \cite{DeWitt}. A single latin index of 
the form $\{i,~j,~ \cdots\}$ indicates the discrete group and Lorentz indices, and the continuous 
space-time variable. For example, a gauge field which would normally be written $A_\mu^a(x)$ becomes 
$\phi_i$. Greek indices of the form $\{\alpha,~\beta,~\cdots\}$ indicate discrete group indices and space-time 
variables.  For example, the Lorentz gauge condition, which would normally be written 
$F^a(x) = \partial^\mu A_\mu^a(x)$ becomes $F_\alpha$. The summation convention is used throughout, 
and is extended to include integration over continuous variables.

\section{The 1PI Effective Action}
\label{1PI}

\subsection{Generalities}

We start from the partition function: 
\begin{eqnarray}
\label{Znoghosts}
&&Z_{1PI}[J]=\int{\cal D}\varphi\; {\rm Det}\left|M_{\alpha\beta}\right| \exp \big[ i\left(I+J^i\varphi_i\right)\big]\; ,~~I = S+S_{gf}\; ,
\end{eqnarray}
where $S$ is the matter and gauge field action. The gauge fixing is set by the action 
$S_{gf}=\frac{1}{2}F_\alpha F^\alpha$. The matrix $M_{\alpha\beta}$ is the Faddeev-Popov operator
\bea
M^\alpha_\beta = \frac{\delta F^\alpha[\varphi]}{\delta\varphi_i}D^i_\beta[\varphi]
\ena
and the functions $D^i_\beta(\varphi)$ represent a complete and independent set of generators of the local gauge transformation.
The only restriction on the gauge fixing condition $F_\alpha[\varphi]$ is that the Faddeev-Popov 
matrix be invertible: we define the ghost propagator by
\bea
M_{\alpha\beta} {\cal G}_{\beta\gamma} = - \delta_{\alpha\gamma}\; ,
\ena
where the $\delta_{\alpha\gamma}$ in this equation is a Kroniker delta. 

It is usually more convenient to work with the effective action, instead of the partition function. 
We start from the generator for connected green functions:
 \bea
 \label{W1}
 W[J] = -i \,{\rm ln}\,Z\; .
 \ena
 The expectation value of the field is obtained from 
 \bea
 \label{exphi}
\phi^i=\langle\varphi^i\rangle=\frac{\delta W}{\delta J_i} \; .
 \ena
 The effective action is defined by the Legendre transform
 \bea
 \label{Gamma1}
 \Gamma[\phi] = W[J]-J_i \phi^i\,;~~ 
 \ena
 Using (\ref{exphi}) we obtain the equations of motion,
\begin{equation}
\label{onshell1}
\frac{\delta \Gamma[\phi]}{\delta \phi^i} = -J_i\; .
\end{equation}

\subsection{Nielsen identities and gauge dependence}

The gauge dependence of the effective action can be explicitly calculated. We consider an infinitesimal change of the gauge condition:
\bea
\label{gt1}
F^\alpha \rightarrow F^\alpha + \delta F^\alpha
\ena 
The goal is to calculate the change produced in the generating functional:
\bea
\delta W = W[F^\alpha+\delta F^\alpha]-W[F^\alpha]\; .
\ena
and to obtain the corresponding change in the effective action by Legendre transforming. 
The calculation can be done in a straightforward way by observing that the change in the action $I$ produced by (\ref{gt1})  can be canceled by a transformation
\bea
\label{shift1}
\varphi\rightarrow \varphi + \delta \varphi\; ,
\ena
which amounts to a shift of integration variables. 
We define $\delta\varphi$ through the equation, 
\bea
\delta(F[\varphi]) = F(\varphi+\delta\varphi)+\delta F(\varphi+\delta\varphi)-F(\varphi) := 0\; .
\ena
Expanding to first order and introducing the notation $F^\alpha_{,i}(\varphi) = \frac{\delta F^\alpha}{\delta\varphi_i}$ we have,
\bea 
\label{gt1b}
F^\alpha_{,i}(\varphi)\delta\varphi_i = -\delta F^\alpha(\varphi)\; .
\ena
The unique solution of (\ref{gt1b}) is,
\bea 
\label{gt2}
\delta \varphi^i= D_\alpha^i[\varphi] {\cal G}^\alpha_\beta [\varphi]\delta F^\beta[\varphi]\; .
\ena
Since this expression has the form of a gauge transformation, the gauge fixed action $S+\frac{1}{2}(F[\varphi])^2$ and the measure ${\cal D}\varphi\; {\rm Det}|M|$ will be invariant under (\ref{gt1}) and (\ref{shift1},\ref{gt2}). Thus, the only contribution to $\delta\Gamma$ will come from the source term. We obtain,
\bea
&&\delta W\big|_{J=const.} = W[F^\alpha+\delta F^\alpha]-W[F^\alpha] = J_i\langle \delta \varphi^i\rangle\nonumber\\[2mm]
&&\delta \Gamma\big|_{J=const.} = \delta W\big|_{J=const.}-J_i \delta \phi^i \nonumber \\[2mm]
&&\delta \phi_i = \langle \delta \varphi _i \rangle + i J_k \langle \delta \varphi^k \varphi_i\rangle-i J_k\langle\delta\varphi^k\rangle \phi_i\; .
\ena
Combining we obtain the Nielsen identity for the 1PI effective action:
\bea
\label{res1PI}
\delta \Gamma\big|_{J=const.} = -i\Gamma_{,i}\Gamma_{,j}\langle (\varphi^i-\phi^i)\delta \varphi^j\rangle
\ena
Thus we find that on the mass shell (defined by $J_i = -\Gamma_{,i}=0$) the effective action is gauge invariant. 

 We note that the variation of the effective action in (\ref{res1PI}) is different from 
the expression found in Eq. (2.14) of \cite{KobesKR2}. Although both results describe the variation 
of the effective action caused by a change in the gauge condition, Eq.~(\ref{res1PI}) is obtained by holding the source 
$J$ constant, and Eq. (2.14) of \cite{KobesKR2} is obtained by holding the mean field $\phi=\phi[J,F]$ 
constant by varying $J$. It is easy to obtain the relationship between these quantities.
We start with the generating function which we consider as a functional of the gauge fixing function and the source:
$W[F,J]$. The variation of the generating function is given by,
\bea
\delta W &&= \frac{\delta W}{\delta F}\Big|_{J=const.} \cdot\delta F +\frac{\delta W}{\delta J}\Big|_{F=const.} \cdot\delta J\nonumber \\[2mm]
&& = \delta W \big|_{J=const.}+\phi\; \delta J
\ena
Subtracting $\delta(J \phi)$ from both sides and using the definition (\ref{Gamma1}) we have,
\bea
\label{deltaG1}
\delta \Gamma = \delta W\big|_{J=const.}-J \delta \phi
\ena
Considering the effective action as a function of the gauge fixing function and the expectation value of the field we have
\bea
\label{deltaG2}
\delta \Gamma &&= \frac{\delta \Gamma}{\delta F}\Big|_{\phi=const.} \cdot\delta F +\frac{\delta \Gamma}{\delta \phi}\Big|_{F=const.} \cdot\delta \phi\nonumber \\
&& = \delta \Gamma \big|_{\phi=const.}-J \delta \phi
\ena
Comparing (\ref{deltaG1}) and (\ref{deltaG2}) we obtain,
\bea
\label{eq:Jphi}
\delta \Gamma \big|_{\phi=const.} = \delta W\big|_{J=const.}  = -\Gamma_{,i}\langle \delta \varphi^i\rangle
\ena
which can be compared with (\ref{res1PI}).
Throughout this paper we will consider variations obtained by holding sources constant. For the 2PI case in particular, this seems to be the more physical choice. The subscripts $J=const$ will be dropped throughout.

\section{The 2PI Effective Action}
\label{2PIall}

\subsection{Generalities}
\label{general}
The 2PI action functional \cite{CornwJT1} can be defined using path integral methods following the technique used to construct the standard 1PI effective action. A bilocal 
source $K$
is needed in addition to the standard local source $J$ to define the partition function $Z[J,K]$. 
The generating function of the connected green function $W[J,K]=-i\ln Z[J,K]$ is 
also a two variable function: 
\begin{equation}
Z[J,K]=e^{iW[J,K]}=\int {\cal D}\varphi e^{i(S[\varphi]+J_i\varphi^i+\frac{1}{2}\varphi_i K^{ij} \varphi_j)}
\end{equation}
The mean field $\phi_i$ and the connected two point function $G_{ij}$ are obtained from $W[J,K]$ as
\begin{equation}
\label{ex-phiG}
\phi^i=\langle\varphi^i\rangle=\frac{\delta W}{\delta J_i}\,;\quad G^{ij}=\langle\varphi^i\varphi^j\rangle-\phi^i\phi^j=i\frac{\delta^2 W}{\delta(iJ_i)\delta(iJ_j)}\; .
\end{equation}   
Differentiating $W[J,K]$ with respect to the bilocal source gives the following 
relation between the mean field and the connected two point function:
\begin{equation}
\label{constraint}
G^{ij}+\phi^i\phi^j=2\frac{\delta W}{\delta K_{ij}}\; .
\end{equation}
The 2PI effective action functional is the Legendre transformation of $W[J,K]$ with 
respect to $J$ and $K$:
\begin{equation}
\label{Gamma2}
\Gamma[\phi,G]=W[J,K]-J_i\phi^i-\frac{1}{2}K_{ij}(\phi^i\phi^j+G^{ij})\; .
\end{equation}   
Using (\ref{ex-phiG}) and (\ref{constraint}) we obtain the following relations:
\begin{equation}
\frac{\delta \Gamma[\phi,G]}{\delta \phi^i}=-J_i-K_{ij}\phi^j\,; ~~ \frac{\delta \Gamma[\phi, G]}{\delta G^{ij}}=-\frac{1}{2}K_{ij}\; .
\label{eq:on-shell}
\end{equation}

The effective action is usually written \cite{CornwJT1} in a more convenient form which has a 
simple diagrammatical interpretation in terms of 2PI diagrams  
\begin{equation}
\label{CJT}
\Gamma[\phi,G]=S_0[\phi]+i\frac{1}{2}\,\rm{Tr}\left\{ \log\left(
G^{-1}\right)+G \left(G_0^{-1}-G^{-1} \right) \right\}+\Phi[\phi,G].
\label{standardform}
\end{equation} 
where $S_0$ is the free part of the action and $G_0$ is the bare two-point function
$\left(-i\delta^2 S_0[\varphi]/\delta \varphi \delta \varphi\right)^{-1}$. The functional 
$\Phi[\phi, G]$ is the sum of all
two-particle-irreducible (2PI) skeleton diagrams with bare vertices and
dressed propagators. In the absence of sources, the difference between the inverse 
dressed and bare propagators is proportional to the one-particle irreducible self energy
\begin{equation}
G^{-1}-G_0^{-1}=2i\frac{\delta \Phi[\phi,G]}{\delta G}\;.
\end{equation}  
This is the usual Schwinger-Dyson equation for the propagator.

The above procedure can be generalized \cite{CornwN1} to construct NPI effective actions. 
For the NPI effective action the skeleton diagrams are calculated using 
dressed propagators as well as dressed N-point proper vertices; the (N+1)-point
vertex is bare.

\subsection{Off-Shell Gauge Dependence} 
\label{nielsen}
To study the gauge dependence of the 2PI effective action we derive the corresponding Nielsen identities. For simplicity, we consider a pure Yang-Mills 
theory. The partition function with a gauge fixing term $F$ has the form,   
\begin{eqnarray}
\label{Z}
&&Z[J,K]=e^{iW[J,K]}\nonumber \\
&&~~~~=\int{\cal D}\varphi\; {\rm Det}\left|\frac{\delta F^\alpha[\varphi]}{\delta\varphi_i}D^i_\beta[\varphi]\right| \exp i\left(S_{YM}+\frac{1}{2}(F[\varphi])^2+J^i\varphi_i+\frac{1}{2}\varphi_iK^{ij}\varphi_j\right)\; .
\end{eqnarray}
 We calculate the variation of the generating functional $W[J,K]$ under the transformations (\ref{gt1}) and (\ref{shift1},\ref{gt2}).
 As in the case of the 1PI effective action, the measure and the gauge fixed action are invariant which means that the only non-zero contribution comes from the source terms.  The sources themselves ($J,~K$) are viewed as the independent variables, and are held constant. We obtain,
\begin{equation}
\label{DW}
\delta W[J,K]= W[F+\delta F]-W[F] =J_i\langle\delta\varphi^i\rangle +\frac{K_{ij}}{2}\left[\langle\delta\varphi^i\varphi^j\rangle+\langle\varphi^i\delta\varphi^j\rangle\right]
\end{equation} 
The variation of the effective action is given by  
\begin{equation}
\label{DGamma}
\delta\Gamma=\delta W -J_i\delta\phi -\frac{K_{ij}}{2}\delta\langle \varphi_i\varphi_j\rangle \;.
\end{equation}
Using the functional definition of expectation values, we obtain
\bea
&&\delta W = \langle T \rangle \nonumber \\[2mm]
&&\delta \phi_i = \langle \delta \varphi_i\rangle +i \langle \varphi_i T \rangle - i \phi_i\langle T \rangle \nonumber \\[2mm]
&&\delta\langle\varphi_i\varphi_j\rangle = \langle\varphi_i \delta\varphi_j+\delta\varphi_i \varphi_j \rangle +i\langle \varphi_i\varphi_j T \rangle - i \langle\varphi_i\varphi_j\rangle \langle T \rangle 
\ena
where we have defined,
\bea
T = J_i\delta \varphi^i + \frac{1}{2}K_{ij}(\varphi^i \delta\varphi^j+\delta\varphi^i \varphi^j)
\ena
Combining these results and using (\ref{eq:on-shell}) we find:
\begin{equation}
\delta\Gamma=-i\left\langle \left(\frac{\delta \Gamma}{\delta\phi^i}\delta\varphi^i+\frac{\delta \Gamma}{\delta G^{ij}}(\delta\varphi^i\xi^j+\delta\varphi^j\xi^i)\right)\left(\frac{\delta \Gamma}{\delta\phi^i}\xi^i+\frac{\delta \Gamma}{\delta G^{ij}}\widetilde{G}^{ij}\right)\right\rangle\; ,
\label{eq:deltaG}
\end{equation}
where $\xi_i=\varphi_i-\phi_i$ and $\widetilde{G}_{ij}=\xi_i\xi_j -G_{ij}$. Using (\ref{eq:on-shell}) we see that the full 2PI effective action is gauge invariant on-shell. Of course, this conclusion is obvious, since the 2PI effective action calculated to all orders is exactly equivalent to the 1PI effective action. One way to see this point is to note that the bilocal source $K$ does not play an active role in (\ref{ex-phiG}). We could set $K$ to zero before differentiating, which shows explicitly that the 2PI $G_{ij}$ is the same as the 1PI $G_{ij}$. \\

Note that the variation of the 2PI effective action at constant mean field and 
propagator can be obtained following the method that was used for the 1PI effective action. We find:
\begin{equation}
(\delta\Gamma)\left.\right|_{\{\phi,\;G\}=const.}=(\delta W)\left.\right|_{\{J,\;K\}=const.}\; .
\end{equation}

\section{Truncation}
\label{Truncation}

All of the preceding calculations are valid when one works with the full effective action (1PI or 2PI). In practice, of course, we never calculate the full effective action: we use an expansion scheme and truncate at some finite order. When we work with a truncated 2PI effective action, the non-perturbative nature of the 2PI formalism gives rise to problems with gauge invariance. In fact, it is easy to see in advance that this problem will occur. The 2PI formalism involves the use of resummed propagators. When calculating at finite orders, one effectively resums a specific class of topologies. Since the Ward identities involve the cancellation of contributions from different topologies, we expect that a resummation that involves only one type of topology will give rise to violations of the Ward identities. In this paper we study the gauge invariance of the effective action. The issue of the Ward Idenities will be left for a future publication \cite{progress}.


We show below that both the 1PI and 2PI effective actions are gauge invariant on shell to arbitrary order in any self-consistent expansion scheme. 

\subsection{1PI}
\label{1PIGammaTrun}

First, consider calculating the full 1PI effective action, {\it i.e.} to all orders in the 
expansion parameter. The mass shell condition is obtained from (\ref{onshell1}) with the source set to zero: $\Gamma_{,i} = 0$. Substituting into (\ref{res1PI}) we have $\delta \Gamma_{1PI} = 0$ which tells us that the full 1PI effective action is gauge invariant on-shell.
We discuss below the truncation of the full 1PI effective action. For definiteness, we consider  a loop expansion. Calculating up to $L$ loops gives the truncated effective action. The full effective action is the sum of the truncated effective action and the remainder:
\bea
\label{Gammatrun}
\Gamma = \Gamma_L+\Gamma_{ex}
\ena
where $\Gamma_L$ is calculated up to $g^{2L-2}$ and $\Gamma_{ex}\sim g^{2L}$.
When using a truncated effective action, the on-shell condition is replaced by an approximate on-shell condition determined from the truncated effective action by,
\bea
\frac{ \delta \Gamma_L[\phi]}{\delta \phi}\Big|_{\phi_{L}^{0}} = 0
\label{truncated eom1}
\ena
Using this approximate on-shell condition we have
\bea
\frac{\delta \Gamma}{\delta \phi}\Big|_{\phi_{L}^{0}} = \frac{\delta \Gamma_{ex}}{\delta \phi}\Big|_{\phi_{L}^{0}} \sim \Gamma_{ex}\sim g^{2L}
\ena
Substituting into (\ref{res1PI}) we obtain,
\bea
\delta(\Gamma_L+\Gamma_{ex}) \sim g^{4L}
\ena
which gives
\bea
\delta \Gamma_L \sim \delta \Gamma_{ex}+ {\cal O}(g^{4L}) \sim  \Gamma_{ex} + {\cal O}(g^{4L}) \sim g^{2L}
\ena
Thus we obtain,
\bea
\label{supp}
\delta \Gamma_L \sim g^2 \Gamma_L
\ena
which shows that the gauge dependence of the on-shell effective action always occurs at higher order than the order of truncation\footnote{The variation of the effective action at constant $\phi$ 
can be evaluated using Eq.(\ref{eq:Jphi}). In this case the
variation of the effective action is linear in the source. Using the approximate on shell condition we obtain  
$$ \left(\delta\Gamma\right)_{\phi=const.}\sim \frac{\delta\Gamma}{\delta\phi}\big|_{\phi_L^{0}}\sim g^{2L}$$
which shows that the gauge variation of total effective action is more weakly suppressed when the expectation value of the field is held constant. Note however that the result for the truncated effective action (\ref{supp}) still holds. }

 It is important to note that the above formal analysis implicitly assumes that the solution of the truncated equation of motion Eq.(\ref{truncated eom1}) remains within the range of validity of the approximation implied by the truncation. In other words, it is possible that $\Gamma_{ex}$, evaluated at the aproximate solution $\phi_0$, contains terms of the same order as $\Gamma_L$. In this case, the truncated effective action, and all physical quantities derived from it,
can be gauge dependent. Such gauge dependence has been observed many times in the literature (the Coleman-Weinberg mechanism \cite{cw,jackiw}, self-consistent dimensional reduction \cite{leivo}, and the one loop plasmon damping rate \cite{KobesKR1,KobesKR2}). The gauge dependence in all of these cases is a manifestation of an inconsistent approximation scheme, since the gauge dependence identities guarantee that physical quantities will be gauge independent when calculated within a self-consistent perturbative procedure.\footnote{Note that a self-consistent perturbative procedure does not necessarily guarantee accuracy. A self-consistent perturbative expansion is one in which an expansion can be carried out in some parameter (such as $\hbar$ or a coupling constant $g$), in such a way that there is no mixing of orders in the expansion. If this self-consistent perturbative expansion exists, the existence of the gauge dependence identities ensures gauge independence of the on-shell effective action order by order, irrespective of the relative magnitude of each term in the expansion. }

\subsection{2PI}
\label{2PIGammaTrun}

Now consider the full 2PI effective action (calculated to all orders). In this case, the conclusion we draw from (\ref{eq:deltaG}) is the same as the conclusion from (\ref{res1PI}): on the mass shell, defined by $\delta \Gamma/\delta\phi^i = \delta \Gamma/\delta G^{ij}=0$, the full 2PI effective action is gauge invariant. As mentioned earlier, this conclusion is obvious, since the 2PI effective action calculated to all orders is exactly equivalent to the 1PI effective action. 
Problems arise when we try to work with a truncated effective action. 
As in the 1PI case, we write the full effective action as the sum of the truncated effective action and the remainder as in (\ref{Gammatrun}). The approximate on-shell condition is determined from the truncated effective action by,
\bea
\label{on-shell-approx}
\frac{\delta \Gamma_{L}}{\delta \phi}\left.\right|_{\phi^0_{L},G^0_{L}} = 0\,;~~ \frac{\delta \Gamma_{L}}{\delta G}\left.\right|_{\phi^0_{L},G^0_{L}} =0.
\ena

Now let's consider performing a perturbative expansion of the type described in section (\ref{1PIGammaTrun}) on (\ref{eq:deltaG}). Using (\ref{on-shell-approx}) we have
\bea
\frac{\delta \Gamma}{\delta \phi}\left.\right|_{\phi^0_{L},G^0_{L}} = \frac{\delta \Gamma_{ex}}{\delta \phi}\left.\right|_{\phi^0_{L},G^0_{L}} \sim \Gamma_{ex}\sim g^{2L}
\label{eq:order}
\ena
Note that in (\ref{eq:order}) we have assumed that the effective action $\Gamma_{ex}$ and its
derivative are of the same order in the expansion.  In fact, it is not clear that this is the case, since the variable that we are differentiating with respect to is a non-perturbative quantity, however, it seems reasonable that differentiation will not increase the order of $\Gamma_{ex}$. We also note that, as in the 1PI case, it is essential that the perturbative procedure is self-consistent, or that the quantity that has been dropped $(\delta\Gamma_{ex}(\phi^0_L,G^0_l))$ is higher order in $g$ than the quantity that is kept $(\delta\Gamma_{L}(\phi^0_L,G^0_l))$. Because of the inherent non-perturbative nature of the 2PI formalism, this possible mixing of orders is more difficult to monitor and control than in the usual 1PI case.
If we assume for the moment that the integrity of the loop expansion is not fundamentally violated, and substitute into (\ref{eq:deltaG}), we obtain as before,
\bea
\delta(\Gamma_L+\Gamma_{ex}) \sim g^{4L}
\ena
which gives
\bea
\delta \Gamma_L \sim \delta \Gamma_{ex}+ {\cal O}(g^{4L}) \sim  \Gamma_{ex} + {\cal O}(g^{4L}) \sim g^{2L}
\ena
Thus we obtain the same result as in the 1PI case:
\bea
\delta \Gamma_L \sim g^2 \Gamma_L
\ena
which shows that the gauge dependence of the on-shell effective action always formally occurs at higher order than the order of truncation, just as for the 1PI case.

\section{The background gauge} 
\label{backg}
In this section we review the background field gauge technique. A more detailed description can be found in \cite{Wein}.  For simplicity, we consider a pure Yang-Mills 
theory.
We write the gauge field as the sum of two pieces:
\begin{equation}
{\rm gauge ~field} = A_\mu + Q_\mu 
\end{equation}
where $A_\mu$ is a background gauge field and $Q_\mu$ is a quantum gauge field which will be the variable of
integration in the functional integral. We add a gauge fixing term (the `background field gauge') that
breaks gauge invariance in $Q_\mu$ but not $A_\mu$. Also, we couple only $Q_\mu$ to sources. This procedure allows us to calculate quantum corrrections and keep explicit gauge invariance in the background field variable.

We define our notation as follows:
\bea
&&[T^a,T^b] = if_{abc}T^c\,;~~~~if^{abc} = (T^b)_{ac}\nonumber\\
&&D_\mu^{ac} = \delta^{ac}\partial_\mu +g f_{abc}A_\mu^b \nonumber \\
&&D_\mu = \partial_\mu-ig A^a_\mu T_a = \partial_\mu-ig A_\mu  \nonumber \\
&& -ig F_{\mu\nu} = [D_\mu,D_\nu] \nonumber \\
&&F_{\mu\nu}^a = \partial_\mu A_\nu^a-\partial_\nu A_\mu^a+gf_{abc}A_\mu^b A_\nu^c 
\ena
Where $T_a$ are the generators of the Lie algebra of the gauge group. 

The partition function is defined as,
\begin{eqnarray}
\label{Z1bkg}
Z[J,K,A] &&= \int [dQ] {\rm det}\left|\frac{\delta G^a}{\delta \alpha^b}\right|{\rm exp}\,\left[i\int d^4x \left({\cal
L}_{YM}(A+Q)-\frac{1}{2\xi}(G^a)^2+J_\mu^a(x) Q^\mu_a(x)\right)\right.\nonumber \\
&&\left. +\frac{i}{2}\int d^4x \int d^4y\,K_{\mu\nu}^{ab}(x,y)Q^\mu_a(x)Q^\nu_b(y)\right]
\end{eqnarray}
We define the generating functional for connected diagrams, and the effective action, in the usual way:
\bea
\label{bkg-Gamma}
&&W[J,K,A] = -i{\rm ln}Z[J,K,A] \nonumber\\
&&\Gamma[\bar Q,\bar G,A] = W[J,K,A]-\int d^4 x J_\mu^a \bar Q^\mu_a-\int d^4x \,d^4y\,\frac{1}{2}K_{\mu\nu}(\bar Q^\mu \bar Q^\nu+\bar G^{\mu\nu}) \nonumber\\
&&
\bar Q_\mu^a = \frac{\delta W}{\delta J^\mu_a}\,;~~\bar G_{\mu\nu} = i\frac{\delta^2 W}{\delta(i J_\mu)\delta(iJ_\nu)}
\ena 
We use the background field gauge condition:
\begin{equation}
\label{bkg}
G^a = \partial_\mu Q^\mu_a+f_{abc}A_\mu^bQ^\mu_c = (D_\mu Q^\mu)^a
\end{equation}
Note that the propagator $\bar G_{\mu\nu}$ should not be confused with the gauge fixing functional $G^a$.
We consider the transformation:
\begin{eqnarray}
\label{bkg-trans}
&&A_\mu^a \rightarrow A_\mu^a+\delta A_\mu^a \,; ~~\delta A_\mu^a = D_\mu^{ab}\alpha_b  \nonumber\\
&&Q_\mu^a\rightarrow Q_\mu^a + \delta Q_\mu^a\,;~~ \delta Q_\mu^a =   f_{abc}Q_\mu^b\alpha^c
\end{eqnarray}
Note that the transformation on $A$ has the form of a gauge transformation. The transformation on $Q$ is just a shift of the integration variable.
It is straightforward to see that (\ref{bkg-trans}) gives,
\bea
\label{bkg-trans2}
&&\delta(A_\mu^a+Q_\mu^a) = \tilde D_\mu^{ab}\alpha_c\,;~~\tilde D_\mu = \partial_\mu-ig(A_\mu+Q_\mu)\nonumber \\
&& \delta (G_a) = f_{abc}G^b \alpha^c
\ena
which means that both the Yang-Mills Lagrangian and the gauge fixing term remain invariant under (\ref{bkg-trans}). 

Combining these results we see that when the transformation (\ref{bkg-trans}) is performed on the generating functional or the effective action, the change that is produced comes only from the source terms. Note that this situation is exactly analogous to the situation we had previously. In sections (\ref{1PI}) and (\ref{2PIall}) we shifted the gauge fixing function and performed a simultaneous shift of the integration variable so that the changes to the generating function and the effective action came only from the source terms. In this case, the shift of the gauge fixing function is generated by a shift in the background field, as shown in (\ref{bkg-trans2}). The calculation of the Nielsen identity follows the same procedure as before.

We can calculate the Nielsen identity for $\Gamma[\bar Q,\bar G,A]$ by following the procedure in section (\ref{nielsen}). We perform the change of variables (\ref{bkg-trans}) and obtain,
\begin{eqnarray}
\delta\Gamma&=&\Gamma[\bar Q,\bar G,A+\delta A]-\Gamma[\bar Q, \bar G, A]\nonumber\\
&=&-i\left\langle \left(\frac{\delta \Gamma}{\delta\phi^i}\delta Q^i+\frac{\delta \Gamma}{\delta G^{ij}}(\delta Q^i\xi^j_A+\delta Q^j\xi^i_A)\right)\left(\frac{\delta \Gamma}{\delta\phi^i}\xi^i_A+\frac{\delta \Gamma}{\delta G^{ij}_A}\widetilde{G}^{ij}_A\right)\right\rangle
\end{eqnarray}
where $\xi^i_A=Q^i-\bar Q^i[A]$, and $\widetilde{G}^{ij}_A=\xi^i_A\xi^j_A-\bar G_{ij}[A]$. As before, we find that the full effective action is gauge invariant on shell, and the truncated effective 
action is gauge invariant to leading order.

\section{Conclusions} 
We have derived the gauge fixing identities for the 2PI effective action, valid for any  gauge fixing function with an invertible Faddeev-Popov matrix. These identities were first derived by Arrizabalaga and Smit \cite{ArrizS1} using a different formalism.
As expected, these identities prove that the 2PI effective action is invariant under infinitesimal gauge variations, on shell, to arbitrary order in any self-consistent expansion scheme. We have
also considered the background field gauge and shown that the effective action is invariant under infinitesimal shifts on the
background field, on shell, to arbitrary order in any expansion scheme.

We note that the derivations of the gauge fixing identities for the 1PI and 2PI formalisms are virtually identical from a mathematical point of view.  This simularity is unexpected, in light of the fact that we expect gauge invarience problems of a completely different nature to arise in the 2PI theory. The 2PI theory is inherently non-perturbative, and involves not just the mean field $\phi_i$ but also the two point function $G_{ij}$, which are {\it a priori} independent, and must be solved for simultaneously. As a consequence, the 2PI theory preferentially resums specific topologies, a procedure that we expect will lead to violations of Ward identities. Of course, since Ward identities reflect the quantum symmetries of the Green functions, such violations are due to truncations and would be absent if exact calculations were possible. 

One possible problem with the truncated 2PI effective action is the fact that the perturbative procedure is not necessarily self-consistent, in the sense discussed in sections 4.1 and 4.2. It is also possible that there is a fundamental problem associated with taking the Legendre transform of a truncated theory. In general, the Legendre transform is defined functionally as the transformation that converts the untruncated generating functional to the untruncated effective action. The 1PI theory is inherently perturbative and thus there is no problem with the Legendre transform for truncated versions of the theory. However, in the case of the 2PI theory, it is possible that problems arise when transforming the truncated theory. In this case, the minimum of the effective potential would not necessarily correspond to the expectation value of the generating functional, and the interpretation of the Ward identities obtained from the effective action would not be straightforward.

\label{conc}


\vspace*{1cm}

\Large

\noindent {\bf Appendix A: BRST transformation} \\

\normalsize
\noindent Our result agrees with the result of \cite{ArrizS1} which was obtained using the BRST method. The basic strategy of the BRST method is to exploit the fact that the gauge-fixed theory still possesses a global symmetry called the BRST symmetry. This symmetry is made explicit by using a representation of the partition function that is different from (\ref{Z}). The gauge fixing term has the form,
\bea
S_{GF} = \int d^4x~ \left(-\bar{c}_\alpha M_{\alpha\beta}c_\beta + B_\alpha V_\alpha -\frac{1}{2}\chi B_\alpha B_\alpha\right) 
\ena
where $c_\alpha$ and $\bar{c}_\alpha$ are the ghost fields, $B_\alpha$ is the auxiliary field, and $V_\alpha$ is the gauge fixing condition. Integration over the ghost fields produces the determinant of the Faddeev-Popov matrix, as in (\ref{Z}), and integration over the auxiliary field produces a gauge fixing term of the form $\frac{1}{2\chi}V_\alpha V_\alpha$. Comparing with (\ref{Z}) we have,
\bea
V_\alpha = \sqrt{\chi}F_\alpha
\ena
To make further progress one must specialize to the covariant gauge: $V_\alpha \rightarrow  \partial^\mu A_\mu^a(x)$.  The gauge fixed action ($S_{YM} + S_{GF}$) is invariant under the BRST transformation: $\delta_{BRST}A_\mu^a = \epsilon (D_\mu c)^a$, $\delta_{BRST} c^a = i\epsilon g (T_a c^a)(T_b c^b)$, $\delta_{BRST} \bar{c}^a = -\epsilon B^a$, and $\delta_{BRST} B^a = 0,$ where $\epsilon$ is an infinitesimal global anti-commuting parameter. One obtains;
\bea
S_{GF} = Q_{BRST} \int d^4 x~ \left(\frac{1}{2} \chi \bar{c}_\alpha B^\alpha -\bar{c}_\alpha V^\alpha\right):=Q_{BRST} \Psi
\ena
where $Q_{BRST}$ is the BRST nilpotent charge defined as $\delta_{BRST}\varphi = \epsilon Q_{BRST}\varphi$. Calculating the variation of the effective action under the BRST transformation one obtains, 
\begin{equation}
\delta\Gamma_{BRST}=\frac{1}{2}\left\langle\delta\Psi Q_{BRST}\left(\frac{\delta \Gamma}{\delta\phi^i}\xi^i+\frac{\delta \Gamma_1}{\delta G^{ij}}\widetilde{G}^{ij}\right)^2\right\rangle\; ,
\end{equation}
with
\begin{equation}
\delta\Psi=-\int d^4x \left(\bar{c}_\alpha\delta V_\alpha[A]-\frac{1}{2}\delta\chi~\bar{c}_\alpha B_\alpha\right)
\end{equation}  
It is straightforward to see that this result is equivalent to ours. We take $V_\alpha = \sqrt{\chi}F_\alpha$ which gives 
\bea
\delta F = \frac{1}{\sqrt{\chi}}\delta V - \frac{1}{2}\frac{\delta \chi}{\chi\sqrt{\chi}} V
\ena
We integrate over the Gaussian $B$ field, and over the ghost fields using 
\begin{equation}
\int {\cal D}c{\cal D}\bar{c} c_i\bar{c}_j e^{-(\bar{c}M c)}=[M^{-1}]_{ij}{\rm det}M\; .
\end{equation}  
We find that the variation 
$\delta\Gamma_{BRST}$ is proportional to our effective action variation $\delta\Gamma$. 
\vskip 1cm



\begin{thebibliography}{99}
\bibitem{Nielsen}
N.K. Nielsen, Nucl. Phys. {\bf B} {\bf 101}, 173 (1975).
\bibitem{KobesKR1}
R. Kobes, G. Kunstatter, Rebhan, Phys. Rev. Lett. {\bf 64}, 2992 (1990). 
\bibitem{KobesKR2}
R. Kobes, G. Kunstatter, A. Rebhan, Nucl. Phys. {\bf B} {\bf 355}, 1 (1991). 
\bibitem{Motto1}
E. Mottola, Proceedings of SEWM 2002, World Scientific publishing (2003), hep-ph/0304279.
\bibitem{ArrizS1}
A. Arrizabalaga, J. Smit, Phys. Rev. {\bf D} {\bf 66}, 065014 (2002).
\bibitem{cw} S. Coleman and E. Weinberg Phys.Rev.D7:1888-1910 (1973).
\bibitem{jackiw} L. Dolan, R. Jackiw, Phys.Rev.D9:2904 (1974).
\bibitem{leivo} G. Kunstatter and H.P. Leivo, Phys.Lett.B166:321(1986); Nucl.Phys.B279:641 (1987).
\bibitem{DeWitt}
B.S. DeWitt, Phys.Rev. {\bf 162}, 1195 (1967); B.S. DeWitt, in {\em Quantum Gravity II}, ed. C.J. Isham, R.Penrose and D.W. Sciama (Oxford Univ. Press, N.Y., 1981) pp. 449-487.
%
\bibitem{CornwJT1}
J.M. Cornwall, R. Jackiw, E. Tomboulis, Phys. Rev. {\bf D 10}, 2428 (1974).
\bibitem{CornwN1}
R.E. Norton and J.M. Cornwall, Ann. Phys. {\bf 91}, 106 (1975).
\bibitem{progress}
Work in progress.
\bibitem{Wein} Steven Weinberg, {\em The Quantum Theory of Fields II}, Cambridge University Press (1996).
\bibitem{Zumin1} 
B. Zumino, J. Math. Phys. {\bf 1}, 1 (1960). 



\end{thebibliography}
\end{document}